\documentclass[preprint2]{aastex}

\begin{document}

\title{An Opportunistic Search for ExtraTerrestrial Intelligence (SETI) with the Murchison Widefield Array}

\author{S.J. Tingay\altaffilmark{1,2},  C. Tremblay\altaffilmark{1}, A. Walsh\altaffilmark{1}, R. Urquhart\altaffilmark{1}}

\altaffiltext{1}{International Centre for Radio Astronomy Research (ICRAR), Curtin University, Bentley, WA 6102, Australia}
\altaffiltext{2}{Istituto di Radio Astronomia, Istituto Nazionale di Astrofisica, Bologna, Italy, 40123}


\begin{abstract}
A spectral line image cube generated from 115 minutes of MWA data that covers a field of view of 400 sq. deg. around the Galactic Centre is used to perform the first Search for ExtraTerrestrial Intelligence (SETI) with the Murchison Widefield Array.  Our work constitutes the first modern SETI experiment at low radio frequencies, here between 103 and 133 MHz, paving the way for large-scale searches with the MWA and, in the future, the low frequency Square Kilometre Array.  Limits of a few hundred mJy/beam for narrow band emission (10 kHz) are derived from our data, across our 400 sq. deg. field of view.  Within this field, 45 exoplanets in 38 planetary systems are known.  We extract spectra at the locations of these systems from our image cube, to place limits on the presence of narrow line emission from these systems.  We then derive minimum isotropic transmitter powers for these exoplanets; a small handful of the closest objects (10s of pc) yield our best limits of order $10^{14}$ W (Equivalent Isotropic Radiated Power: EIRP).  These limits lie above the highest power directional transmitters near these frequencies currently operational on Earth.  A SETI experiment with the MWA covering the full accessible sky and its full frequency range would require approximately one month of observing time.  The MWA frequency range, its Southern Hemisphere location on an extraordinarily radio quiet site, its very large field of view, and its high sensitivity make it a unique facility for SETI.
\end{abstract} 


\section{INTRODUCTION}
The first modern Search for ExtraTerrestrial Intelligence (SETI) was undertaken at radio wavelengths in 1960 at Green Bank, West Virginia \citep{dra08}, targeting two stars, Tau Ceti and Epsilon Eridani at frequencies near the 21-cm line of neutral hydrogen.  

In the decades since, SETI programs have continued to be undertaken at radio wavelengths, on the basis that highly sensitive radio telescopes exist for astronomy, the radio band is the cornerstone of communications technologies on Earth, and it could be reasonably assumed that a similar technology path has been taken by extraterrestrial civilisations.  In 1985, the Million-channel ExtraTerrestrial Assay (META) was established at Harvard University \citep{lei97}, also near 1.4 GHz.  META was upgraded to the Billion-channel ExtraTerrestrial Assay (BETA) in 1995, in the extended frequency range of 1.4 - 1.7 GHz.  The SERENDIP (Search for Extraterrestrial Radio Emissions from Nearby Developed Intelligent Populations) program was established in 1978 and has evolved considerably since then \citep{wer01}.  The SERENDIP program also gave rise to projects such as SETI@home \citep{kor09} and Southern SERENDIP \citep{sto00}, a survey conducted using the Parkes radio telescope. A novel targeted search using Very Long Baseline Interferometry observations of the Gliese 581 star system is described by \citet{ram12}.  The Allen Telescope Array (ATA) has been used extensively for a range of SETI experiments over the last decade \citep{wel09}.  See \citet{gar14} for a recent review of aspects of SETI experiments at radio wavelengths.  For general reviews, see \citet{sie15,tar03}.

The SETI experiments conducted at radio wavelengths to date have generally focused on the 1.4 - 1.7 GHz range, the so-called ``water hole" between the prominent radio spectral lines due to neutral hydrogen (H) and hydroxyl (OH).  However, many other radio frequencies are also viable for SETI experiments.  One radio frequency range that has opened up in recent years is in the tens to hundreds of MHz.  Powerful multi-purpose, next-generation low frequency radio telescopes such as (LOFAR: \citealt{van14}) and the Murchison Widefield Array (MWA: \citealt{tin13,lon09}) are precursors and pathfinders for the billion dollar Square Kilometre Array (SKA: \citealt{dew09}) over the next decade.  A key science program for the SKA is the ``cradle of life" \citep{hoa15}, including comprehensive and ambitious SETI experiments \citep{sie15}.  \citet{loe07} discuss the prospects for SETI experiments using facilities such as the MWA, LOFAR, and the low frequency SKA, instruments that are primarily designed to search for the redshifted neutral hydrogen line from the Epoch of Reionisation.  \citet{rah15} discusses the potential for utilising microlensing for SETI programs at radio wavelengths, with particular reference to the MWA and SKA.  LOFAR, operating in the ranges of 30 - 80 MHz and 120 - 240 MHz, launched a SETI project in 2010\footnote{https://www.astron.nl/about-astron/press-public/news/lofar-opens-low-frequency-universe-and-starts-new-seti-search/lofar-o}, although no results have thus far been published.

The MWA operates in the frequency range of 80 - 300 MHz on an exceptionally radio quiet site in Western Australia.  The high sensitivity of the MWA, its radio quiet location \citep{2015PASA...32....8O}, its frequency range, its access to the Southern Hemisphere, and its exceptionally large field of view (hundreds of square degrees) make it a unique facility for exploratory SETI experiments.  As \citet{gar14} points out, the emergence of new radio telescopes with very large fields of view opens up new areas of parameter space for SETI experiments.  For example, within a single $>$500 sq. deg. field of view typical for the MWA, on average tens of stellar systems within 50 lightyears will be accessible \citep{hip97}.  Based on recent results showing that planets are the norm rather than the exception, for example on average $1.0\pm0.1$ planets per M dwarf star in our Galaxy \citep{swi13}, one would therefore expect dozens of nearby (within 50 lightyears) planets in a single MWA field of view and far greater numbers of more distant planets.  The MWA field of view therefore results in a significant multiplex advantage that can be exploited for SETI experiments.

While facilities such as the MWA, LOFAR, and the SKA will open up unprecedented parameter space for new SETI experiments, it is worth noting that the very first consideration of low radio frequency SETI came approximately 60 years before the first modern experiments described in \citet{dra08}, at the dawn of radio communications.  In the late 1800s and early 1900s, Guglielmo Marconi\footnote{Reported extensively in the media at the time, for example on page 3 of the New York Tribune, September 2, 1921: http:\/\/chroniclingamerica.loc.gov\/lccn/sn83030214\/1921-09-02\/ed-1\/seq-3\/} and Nicola Tesla \footnote{As described by Corum \& Corum (1996): http:\/\/www.teslasociety.com\/mars.pdf} believed that radio waves could be used to communicate with civilisations on Mars (the widespread belief in the existence of Martian canals persisted at the time) and both claimed to have detected potential signals from that planet.  In 1924, an experiment to listen for signals was organised by the US Navy during the opposition of Mars that year, coordinated with a planned cessation of terrestrial radio broadcasts\footnote{http:\/\/www.lettersofnote.com\/2009\/11\/prepare-for-contact.html}.  These very first low radio frequency experiments returned null results.

In this Letter, we present a first, and opportunistic, SETI pilot experiment with the MWA, in the frequency range 103 - 133 MHz, placing limits on narrow band radio emission toward 38 known planetary systems.  The experiment is opportunistic in the sense that the observations were undertaken for a spectral line survey of the Galactic Plane that is ongoing; utility of the data for a SETI experiment was realised post-observation.  We use this pilot study to motivate a deeper and significantly larger SETI experiment with the MWA, that could use the full 80 - 300 MHz frequency range and survey the entire southern sky (majority of the Milky Way) visible from Western Australia.
 
The field of SETI has recently received a substantial boost, with the ``Breakthrough Listen" project recently initiated\footnote{http://www.breakthroughinitiatives.org/}.  Novel and diverse SETI experiments that sit on the path to utilisation of the SKA for SETI, such as described here, are likely to be useful contributors to such initiatives.

\section{OBSERVATIONS AND DATA ANALYSIS}
The MWA system is described in \cite{tin13}, so we provide only a brief summary of the salient aspects of the observations.  Observations with the MWA took place on 25 July 2014. Dual-polarisation data were obtained in a 30.72 MHz contiguous bandwidth (consisting of 24$\times$1.28 MHz contiguous coarse channels), centered at 119.7 MHz, with 10 kHz frequency resolution (thus 128 fine channels per 1.28 MHz coarse channel). Observations took place in five minute segments over a total of 115 minutes, with pointing coordinates of RA=17h45m40.036s and Dec=-29$^{\circ}$00′28.17′′ and a field of view (primary beam) FWHM of $\sim30^{\circ}$ at a resolution (synthesized beam) of 3\arcmin.  However, only 400 square degrees were imaged and searched, in order to make this first search a manageable size.

The edges of each coarse channel suffer from aliasing, requiring a number of fine channels on each coarse channel edge to be flagged.  This resulted in approximately 78\% of the 3072 fine channels being imaged. The central fine channel of each coarse channel was flagged, as they contain the DC component of the filterbank.  Automated flagging of radio frequency interference (RFI) was performed using AO Flagger \citep{off12} and a small number of fine channels were manually removed.  In total, less than 0.5\% of data were removed due to RFI, consistent with previous examinations of the RFI environment of the MWA \citep{2015PASA...32....8O}.  As described in \citet{tin15}, AO Flagger only removes the strongest RFI, well above any limits on SETI signals discussed below.  We also note that the RFI removed all corresponds to known terrestrial transmitter frequencies (generally FM radio and digital TV broadcast frequencies).

Calibration and imaging proceeded using the techniques described in \citet{hur14}.   This includes using WS Clean \citep{off14}, using w-stacking to take into account the w-terms, to clean and image each individual fine channel with Briggs weighting set to $-1$, a compromise between natural and uniform weighting to balance signal to noise and image resolution.  The 100 individual fine channels from each coarse channel are imaged using 5.7 pixels per synthesised beam full width at half maximum and were combined to produce a single data cube which was then converted into a MIRIAD \citep{1995ASPC...77..433S} file format.  In MIRIAD, all of the five-minute observations were time-averaged together to form a single data cube, representing an effective 82 minutes of integration time.

A continuum-subtracted image cube was produced, removing 99\% of continuum emission using custom software. The field of view imaged for the cube was 400 sq. deg, covering the right ascension range 17h03m24s to 18h28m06s and the declination range -18$^{\circ}$34′48′′to -38$^{\circ}$30′44′′. The RMS noise levels in the continuum-subtracted cube (Table 1) were in line with theoretical noise estimates, calculated from the system description in \citet{tin13}.

Although an analytic model for a dipole over a ground screen has been used previously for primary beam correction with MWA data \citet{hur14}, we made no correction for the beam model in the creation of these data cubes, since this experiment is primary a detection experiment. To estimate the variation in expected flux densities, the data from the Molonglo Reference Catalogue (MRC; \citet{lar81}), NRAO VLA Sky Survey (VSS; \citet{con98}), and Parkes-MIT-NRAO survey (PMN; \citet{gri94}) were compared to the flux density recovered for extragalactic sources in a non-continuum subtracted image.  This showed an error in the flux densities between 7 percent and 38 percent depending on angular distance from the phase centre.  In order to be conservative, the upper limits listed in Table 1 include a 40\% error in the flux density scale.

A search of the field for exoplanets, based on the Kepler catalogue \citep{kep13}, returned 38 known planetary systems containing 45 exoplanets. The systems are listed in Table 1 and are shown in relation to the MWA field of view in Figure 1.

The MWA data cube was searched at the locations of each of these exoplanet systems and spectra were extracted from these locations. No significant narrow band signals were detected above a 5$\sigma$ level in any of these spectra. Table 1 lists the RMS from the spectra at each of the exoplanet system locations and the corresponding 1$\sigma$ limits on the inferred isotropic transmitter power required at the distance of the exoplanet system (ranging between 10$^{14}$ W and 10$^{20}$ W). Figure 2 shows example spectra extracted from the MWA data cube, representing the four closest stars in Table 1: GJ 667; HD 156846; HD 164604; and HD 169830.

\begin{figure}[ht]
\centering
\includegraphics[width=6cm,angle=0]{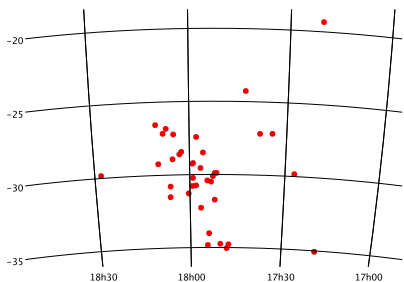}
\caption{Distribution of the 38 known exoplanet systems in the MWA field of view.}
\end{figure}

GJ 667, at 6.8 pc, is the closest star in Table 1. The closest recorded star to Earth in our field is 36 Ophiuchi, at a distance of 5.9 pc, which is not significantly closer than GJ 667 when considering the calculation of limits on transmitter power in Table 1. For the typical spectral RMS in Table 1, four to five hundred mJy/beam, the corresponding limit for the nearest extrasolar star to the Earth (proxima Centauri, 1.3 pc - not in our field) would be approximately 10$^{13}$ W.

\begin{table*}[ht]
\centering
\small
  \begin{tabular}{c l r c c c c} \hline 
\# System & RA & Dec & RMS (mJy) & Dist. (pc) & P ($10^{13}$ W) \\ 
GJ 667 C & 259.7451 & -34.9968 & 529 & 6.8 & $<$4   \\ 
HD 156846 & 260.1429 & -19.3337 & 461 & 49 & $<$185   \\ 
HD 164604 & 270.7789 & -28.5606 & 488 & 38 & $<$117   \\ 
HD 169830 & 276.9562 & -29.8169 & 453 & 36 & $<$98   \\ 
MOA 2007-BLG-192L & 272.0158 & -27.1501 & 457 & 1000 & $<$8$\times10^{4}$   \\ 
MOA 2007-BLG-400L & 272.4249 & -29.2242 & 470 & 5800 & $<$3$\times10^{6}$   \\ 
MOA 2008-BLG-310L & 268.5605 & -34.7781 & 461 & 6000 & $<$3$\times10^{6}$   \\ 
MOA 2008-BLG-379L & 269.7060 & -30.1969 & 443 & 3300 & $<$8$\times10^{5}$   \\ 
MOA 2009-BLG-266L & 267.0081 & -35.0054 & 454 & 3040 & $<$7$\times10^{5}$   \\ 
MOA 2009-BLG-319L & 271.7422 & -26.8197 & 465 & 6100 & $<$3$\times10^{6}$   \\ 
MOA 2009-BLG-387L & 268.4616 & -33.9903 & 448 & 5690 & $<$2$\times10^{6}$   \\ 
MOA 2010-BLG-073L & 272.5473 & -26.5229 & 489 & 2800 & $<$6$\times10^{5}$   \\ 
MOA 2010-BLG-328L & 269.4963 & -30.7152 & 467 & 810 & $<$5$\times10^{4}$   \\ 
MOA 2010-BLG-477L & 271.5310 & -31.4545 & 460 & 2300 & $<$4$\times10^{5}$   \\ 
MOA 2011-BLG-262L & 270.0978 & -31.2453 & 474 & 7000 & $<$4$\times10^{6}$   \\ 
MOA 2011-BLG-293L & 268.9140 & -28.4768 & 480 & 7720 & $<$5$\times10^{6}$   \\ 
MOA 2011-BLG-322L & 271.2233 & -27.2209 & 469 & 7560 & $<$4$\times10^{6}$   \\ 
MOA-bin-1L & 261.7925 & -29.7940 & 452 & 5100 & $<$2$\times10^{6}$   \\ 
OGLE 2003-BLG-235L & 271.3181 & -28.8950 & 478 & 5200 & $<$2$\times10^{6}$   \\ 
OGLE 2005-BLG-169L & 271.5222 & -30.7326 & 469 & 2700 & $<$6$\times10^{5}$   \\ 
OGLE 2005-BLG-390L & 268.5800 & -30.3773 & 474 & 6600 & $<$3$\times10^{6}$   \\ 
OGLE 2005-BLG-71L & 267.5407 & -34.6732 & 456 & 3200 & $<$8$\times10^{5}$   \\ 
OGLE 2006-BLG-109L & 268.1438 & -30.0878 & 469 & 1510 & $<$2$\times10^{5}$   \\ 
OGLE 2007-BLG-368L & 269.1082 & -32.2374 & 464 & 5900 & $<$3$\times10^{6}$   \\ 
OGLE 2008-BLG-355L & 269.7867 & -30.7595 & 476 & 6800 & $<$4$\times10^{6}$   \\ 
OGLE 2008-BLG-92L & 266.8726 & -34.7266 & 455 & 8100 & $<$5$\times10^{6}$   \\ 
OGLE 2011-BLG-251L & 264.5591 & -27.1361 & 469 & 2570 & $<$5$\times10^{5}$   \\ 
OGLE 2011-BLG-265L & 269.4488 & -27.3945 & 506 & 4380 & $<$2$\times10^{6}$   \\ 
OGLE 2012-BLG-26L & 263.5779 & -27.1428 & 453 & 4080 & $<$1$\times10^{6}$   \\ 
OGLE 2012-BLG-358L & 265.6949 & -24.2610 & 443 & 1760 & $<$2$\times10^{5}$   \\ 
OGLE 2012-BLG-406L & 268.3257 & -30.4712 & 472 & 4970 & $<$2$\times10^{6}$   \\ 
OGLE 2013-BLG-102L & 268.0295 & -31.6906 & 457 & 3040 & $<$7$\times10^{5}$   \\ 
OGLE 2013-BLG-341L B & 268.0312 & -29.8461 & 475 & 1161 & $<$1$\times10^{5}$   \\ 
OGLE 2014-BLG-124L & 270.6217 & -28.3963 & 470 & 4100 & $<$1$\times10^{6}$   \\ 
OGLE-TR-056 & 269.1480 & -29.5392 & 478 & 1500 & $<$2$\times10^{5}$   \\ 
OGLE-TR-10 & 267.8677 & -29.8764 & 475 & 1500 & $<$2$\times10^{5}$   \\ 
SWEEPS-11 & 269.7583 & -29.1982 & 470 & 8500 & $<$6$\times10^{6}$   \\ 
SWEEPS-4 & 269.7247 & -29.1891 & 473 & 8500 & $<$6$\times10^{6}$   \\ \hline
 \end{tabular}
   \caption{The 38 known exoplanet systems in the MWA field of view.  Column 1 - Exoplanet system; Column 2 - right ascension (deg); Column 3 - declination (deg); Column 4 - RMS (mJy/beam); Column 5 - distance (pc); and Column 6 - upper limit on isotropic transmitter power in units of $10^{13}$ W}
\end{table*}

\begin{figure*}[ht]
\centering
\includegraphics[width=12cm,angle=270]{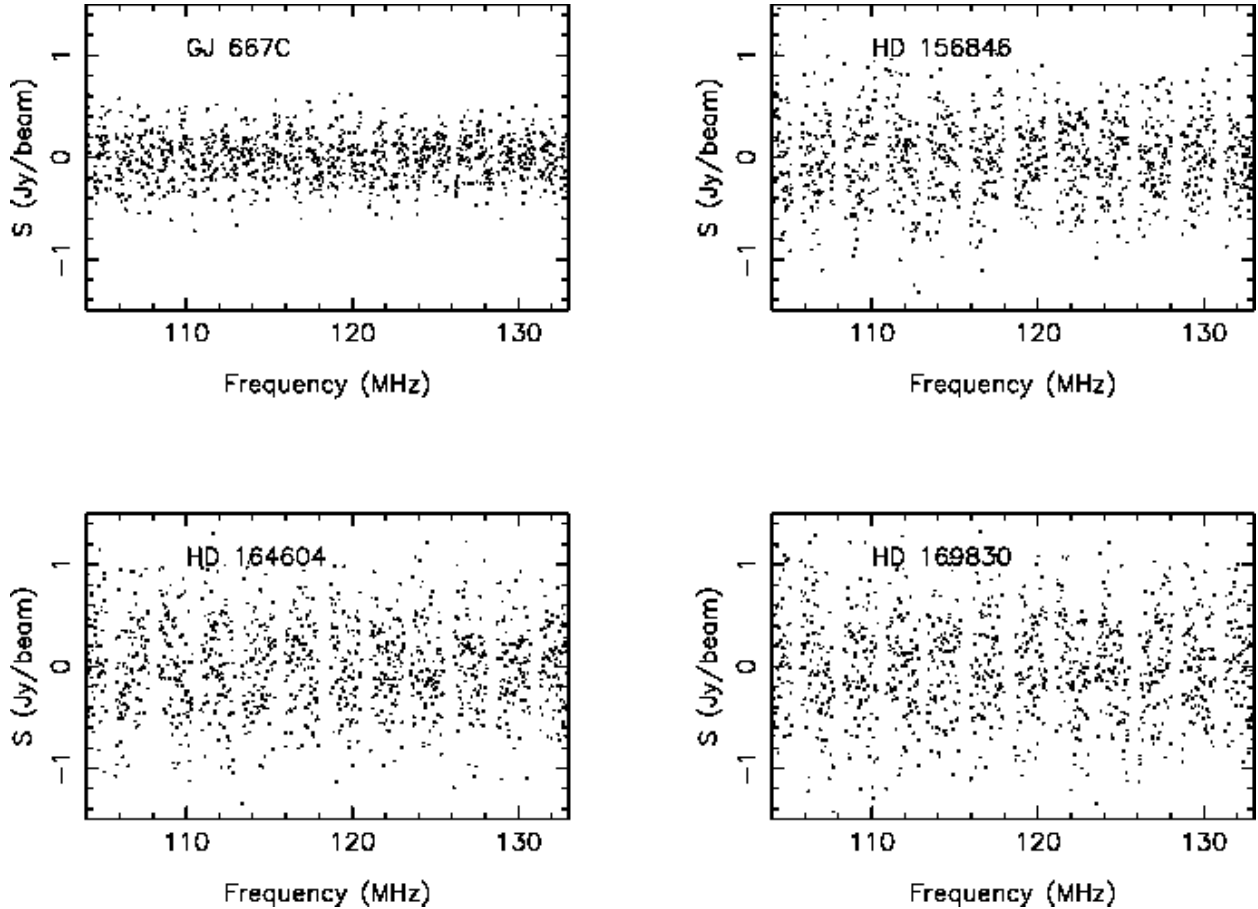}
\caption{MWA spectra for the four closest exoplanet systems listed in Table 1.  Areas of flagged fine channels on coarse channel edges are evident as gaps in the spectra.}
\end{figure*}

\section{DISCUSSION AND CONCLUSION}
The great majority of exoplanet systems listed in Table 1 are at large distances, yielded from microlensing experiments toward the Galactic Centre, meaning that the observational limits on detectable transmitter power from the MWA are very high, inferred isotropic powers of $10^{17} - 10^{19}$ W.  Even if a directional transmitting antenna is assumed, with a gain similar to low frequency over-the-horizon radar transmitters on Earth, the limits on transmitter power are only reduced by factors of order 100.

A small handful of exoplanet systems in Table 1 are close enough that the inferred isotropic transmitter powers are of order $10^{13} - 10^{14}$ W.  These are still very large in terms of transmitters on Earth.  The highest power low frequency transmitters on Earth are the over-the-horizon (OTH) backscatter radars used for military surveillance; these typically operate in the 5 - 30 MHz range and have transmitter powers of order 1 MW.  For example, the Jindalee Operation Radar Network (JORN) in Australia has a transmitter power of 560 kW \citep{col00} and similar installations in the US, such as the  in the AN/FPS-118 OTH-Backscatter radar\footnote{http://www.globalsecurity.org/wmd/systems/an-fps-118.htm}, have transmitter powers of 1 MW (but can range up to 10 MW).  In the latter case of the US system, the Effective Radiated Power is 100 MW, still a factor of $\sim10^{5}$ below the limits for the nearest exoplanet systems in Table 1.  Even the addition of the signals from the ensemble of global array of OTH radars fall well below our limits.  \cite{loe07} summarise other Earth-based transmitter characteristics relevant to low frequency telescopes.

The most powerful transmission ever broadcast deliberately into space was the Arecibo message, directed as a purposeful communication at the globular cluster M13, in a 10 Hz bandwidth at 2380 MHz \citep{ca75}.  This transmission had an equivalent isotropic transmission of $20\times10^{9}$ W.  Taking the narrow bandwidth into account (and ignoring the large difference in frequency), this transmission once again falls below the limits calculated in Table 1.

In the MWA frequency range, \citet{mck13} previously estimated the Equivalent Isotropic Power of FM radio transmissions from the Earth to be 77 MW.  This estimate was made by measuring the amount of stray FM radio signal reflected off the Earth's Moon.  Again, this isotropic power is well below the limits in Table 1.

These projections of Earth-based technologies of course discount the possibility that higher power and/or more highly directive antenna technologies are utilised by advanced extraterrestrial civilisations for communications or remote sensing applications.

While the inferred transmitter powers in Table 1 are high compared to the most powerful low frequency transmitters on Earth, this study has nonetheless provided the most comprehensive search for narrow band transmissions from exoplanets in this frequency range.  Due to the southern, RFI free location of the MWA, its operational frequency range, and its wide field of view, the MWA provides a unique capability for future SETI projects.

This experiment examined one field of view of 400 sq. deg. in a 30.72 MHz frequency band.  To perform a SETI experiment to the same depth as achieved here, but over the full MWA frequency range (80 - 300 MHz), and over the full accessible sky from Western Australia, would require approximately one month of observing time.  This is an entirely feasible goal for the near future (three times deeper would require of order a year of observing).  Moreover, to relieve the restriction of 10 kHz frequency resolution present in the current experiment, it is possible to record voltage data from the MWA and reconstruct coherent beams at far higher spectral resolution to target individual exoplanet systems \cite{tre14}.  For example, generating 1 Hz channels from coherent beams across the full array would yield a factor of 100 improvement in sensitivity (assuming a 1 Hz transmission bandwidth), compared to the current experiment.  Such a mode could be run communally with the large-scale survey described above, for a selected list of target systems.

The current experiment and the capabilities of the MWA provide a clear path to the far greater capabilities of the low frequency component of the Square Kilometre Array, which will be built at the same location as the MWA and have a spectral sensitivity some tens of times greater than the MWA.  The radio quiet nature of the MWA/SKA site in enabling SETI experiments at low frequencies (especially through the FM band), as demonstrated here, bodes well for SETI experiments with the SKA.

\section{Acknowledgements}
This scientific work makes use of the Murchison Radio-astronomy Observatory, operated by CSIRO. We acknowledge the Wajarri Yamatji people as the traditional owners of the Observatory site.  We thank Ian Morrison and Andrew Siemion for discussions regarding a draft of this Letter.  Support for the operation of the MWA is provided by the Australian Government Department of Industry and Science and Department of Education (National Collaborative Research Infrastructure Strategy: NCRIS), under a contract to Curtin University administered by Astronomy Australia Limited. We acknowledge the iVEC Petabyte Data Store and the Initiative in Innovative Computing and the CUDA Center for Excellence sponsored by NVIDIA at Harvard University. 

{\it Facility:} \facility{MWA}.

\end{document}